\documentclass{jfm}
\usepackage{graphicx}
\usepackage{soul}
\usepackage{epstopdf, epsfig}
\usepackage{amsmath,tikz,commath,appendix,color} 
\usepackage[normalem]{ulem}

\shorttitle{Viscous propulsion in active transversely-isotropic media}
\shortauthor{G. Cupples, R. J. Dyson and D. J. Smith}

\title{Viscous propulsion in active transversely-isotropic media \\
CORRIGENDUM}

\author{G. Cupples,
	R. J. Dyson
 \and  D. J. Smith
  \corresp{\email{D.J.Smith@bham.ac.uk}}}

\affiliation{School of Mathematics, University of Birmingham, B15 2TT, U.K.
}

\begin{document}

\maketitle

\section{Introduction}

Small organisms swimming at very low Reynolds numbers, for example spermatozoa in cervical mucus, cannot propel themselves by utilising the inertia of the surrounding fluid; time-reversible kinematics result in zero net displacement for the small body. G.I. Taylor's pioneering study presented the first model of zero-Reynolds-number swimming where time-reversal symmetry is broken by the wave direction \citep{taylor1951swimming}. This model was formulated as the far-field Stokes flow produced by a swimming motion given by a small amplitude sinusoidal wave, and the mean rate of working was calculated as a measure of the energetic cost of swimming. Our recent study, `Viscous propulsion in active transversely-isotropic media' \citep{cupples2017viscous} adapted Taylor's model to account for fibre-reinforced media, similar in nature to the glycofilament structure of cervical mucus, through the transversely-isotropic constitutive equations of \citeauthor{ericksen1960ti} (\citeyear{ericksen1960ti}); we now detail a corrigendum, which in particular shows the importance of fibre orientation for both passive and active fluid cases.

Our paper consisted of calculating the mean swimming velocity and energy dissipation of an infinite waving sheet in a transversely-isotropic fluid in 2D, extending the classical Taylor's swimming sheet model to include anisotropic effects and active rheology. A surprising conclusion was that fibre orientation only affected swimming velocity in the active case. However a recent study by \citet{shi2017swimming} investigated microscopic propulsion in nematic liquid crystals and found that in a common limit (passive, zero elasticity, zero shear viscosity and small extensional viscosity) the models disagreed, with their study finding an angle-dependence in swimming speed. Here we find that this discrepancy is due to missed terms in the solution of the governing equation in \citep{cupples2017viscous}. These terms are relevant to both the passive and active cases, and qualitatively change the conclusions.

The analysis involves a perturbation expansion in the small parameter $\varepsilon=k^*b^*$, where $b^*$ is the amplitude and $k^*$ the wavenumber. The leading order solution at $\mathcal{O}(\varepsilon)$ is unchanged from \cite{cupples2017viscous}, and we here discuss a correction to the $\mathcal{O}(\varepsilon^2)$ solution which determines the swimming velocity. First the passive transversely-isotropic fluid case is discussed ($\mu_1=0$) in section \ref{no_mu1_sol}, which is shown to be consistent with \citet{shi2017swimming} in a common limit; the mean swimming velocity is recalculated and presented for a wide range of anisotropic extensional and shear viscosities in section \ref{no_mu1_Results}. After this, a solution to the active case is considered in section \ref{mu1_sol}, where a spatially averaged swimming velocity is calculated and discussed.

\subsection{Equation formulation}
The full system of equations is derived from the dimensionless Navier-Stokes equations, at zero Reynolds number, along with \citeauthor{ericksen1960ti}'s \citeyear{ericksen1960ti} constitutive equation for a transversely isotropic fluid (equations (2.1)--(2.3) in \citeauthor{cupples2017viscous} \citeyear{cupples2017viscous}. A stream function $\psi$, satisfying incompressibility, and an equation governing the perturbation to the fibre orientation $\theta$ around a uniform initial fibre angle $\phi$ (equation (2.5) in \citeauthor{cupples2017viscous} \citeyear{cupples2017viscous}) complete the model. At $\mathcal{O}(\varepsilon^2)$ the system of partial differential equations is
\begin{eqnarray}
& &\left(1+\frac{\mu_2}{4}\sin^22\phi+\mu_3\right)\nabla^4\psi_1-\mu_1\left(2\sin2\phi\dmd{\theta_1}{2}{x}{}{y}{}+\cos2\phi\left(\dpd[2]{\theta_1}{x}-\dpd[2]{\theta_1}{y}\right)\right) \nonumber  \\
& & \hspace{1.5cm} +\mu_2\left(\cos4\phi\dmd{\psi_1}{4}{x}{2}{y}{2}+\frac{\sin4\phi}{2}\left(\dmd{\psi_1}{4}{x}{}{y}{3}-\dmd{\psi_1}{4}{x}{3}{y}{}\right)\right)=F(\psi_0,\theta_0),\label{gov1} \\
& & \hspace{2.25cm}\dpd{\theta_1}{t}+\sin^2\phi\dpd[2]{\psi_1}{y}+\sin2\phi\dmd{\psi_1}{2}{x}{}{y}{}+\cos^2\phi\dpd[2]{\psi_1}{x}=G(\psi_0,\theta_0),\label{gov2}
\end{eqnarray}
where $F$ and $G$ are known functions of the $\mathcal{O}(\varepsilon)$ solutions and are given in appendix \ref{AppA}. This is stated in full in equation (C 1) of appendix C in \cite{cupples2017viscous}. These functions involve terms proportional to $\cos^2(x-t)$,  $\sin^2(x-t)$ and $\sin(x-t)\cos(x-t)$ with coefficients in terms of the anisotropic parameters.

In section \ref{no_mu1_sol} we take $\mu_1=0$, which we refer to as the `passive fluid' case, and solve the resulting system to determine the mean swimming velocity. The steps in this calculation are elucidated in more detail in order to highlight how to correct the solution.  In section \ref{mu1_sol} we reconsider the active case for nonzero $\mu_1$.

\section{Mean swimming velocity in a passive fluid} \label{no_mu1_sol}

For a passive fluid, {\it i.e} when $\mu_1=0$, the system of equations at $\mathcal{O}(\varepsilon^2)$, \eqref{gov1} and \eqref{gov2}, become
\begin{eqnarray}
& &\left(1+\frac{\mu_2}{4}\sin^22\phi+\mu_3\right)\nabla^4\psi_1 \nonumber \\
& & \quad +\mu_2\left(\cos4\phi\dmd{\psi_1}{4}{x}{2}{y}{2}+\frac{\sin4\phi}{2}\left(\dmd{\psi_1}{4}{x}{}{y}{3}-\dmd{\psi_1}{4}{x}{3}{y}{}\right)\right)=F(\psi_0,\theta_0), \label{psi} \\
& &\dpd{\theta_1}{t}+\sin^2\phi\dpd[2]{\psi_1}{y}+\sin2\phi\dmd{\psi_1}{2}{x}{}{y}{}+\cos^2\phi\dpd[2]{\psi_1}{x}=G(\psi_0,\theta_0),\label{theta}
\end{eqnarray}
along with boundary conditions (given as (3.31)-(3.32) in the original paper)
\begingroup
\addtolength{\jot}{0.7em}
\begin{eqnarray}
\dpd{{\psi_1}}{y}\bigg|_{y=0}&=&\frac{1}{2}\left((\alpha_1\alpha_2-\beta_1\beta_2)(1-\cos 2(x-t))-(\alpha_1\beta_2-\alpha_2\beta_1)\sin 2(x-t)\right) ,\label{ep2bc1}\\
\dpd{{\psi_1}}{x}\bigg|_{y=0}&=&0.\label{ep2bc2}
\end{eqnarray}
\endgroup
Upon substitution of $\psi_0$ and $\theta_0$ into \eqref{extraF} and \eqref{extraG}, the inhomogeneous terms (when $\mu_1=0$) take the form
\begin{eqnarray}
F&=&m_1\cos^2(x-t)+m_2\cos(x-t)\sin(x-t)+m_3\sin^2(x-t),\nonumber \\
&=&\frac{m_1+m_3}{2}+\frac{m_1-m_3}{2}\cos 2(x-t)+\frac{m_2}{2}\sin 2(x-t),\\
G&=&\frac{n_1+n_3}{2}+\frac{n_1-n_3}{2}\cos 2(x-t)+\frac{n_2}{2}\sin 2(x-t),
\end{eqnarray}
where $m_j=\sum_{k=1}^{10}M_j^{(k)}\exp(\gamma_k y)$ and $n_j=\sum_{k=1}^{10}N_j^{(k)}\exp(\gamma_k y)$ for $j=1,2,3$. There are ten possibilities for $\gamma_k$, resulting from combinations of $\psi_0$ and $\theta_0$, which are
\begin{eqnarray}
\gamma_1=&2\lambda_1,\,\gamma_2=2\lambda_2,\,\gamma_3=2\lambda_3,\,\gamma_4=2\lambda_4,\,\gamma_5=\lambda_1+\lambda_2,\,\gamma_6=\lambda_1+\lambda_3, \nonumber \\
&\gamma_7=\lambda_1+\lambda_4,\,\gamma_8=\lambda_2+\lambda_3,\,\gamma_9=\lambda_2+\lambda_4,\,\gamma_{10}=\lambda_3+\lambda_4, \label{rootop}
\end{eqnarray}
where $\lambda_j$ are determined as part of the leading order solution. Since all $\lambda_j$ have negative real part, to satisfy the far-field condition at leading order, the real part of these exponentials will always be negative. By setting $\mu_1=0$ the governing equations \eqref{psi} and \eqref{theta} decouple; since we are interested in the mean swimming velocity we focus on the solution to \eqref{psi} only.

\subsection{Corrected solution} \label{no_mu1_method}
The first step we take is to note that $x$ and $t$ only appear together as $x-t$ and so we make the substitution $z=x-t$; in what follows we will be precise regarding which variable we are averaging over as the active case is not $t$-periodic in general.

Equation \eqref{psi} becomes
\begin{eqnarray}
\left(1+\frac{\mu_2}{4}\sin^22\phi+\mu_3\right)&\nabla^4&\Psi_1 \nonumber \\
& & \hspace{-2.8cm} +\mu_2\left(2\cos4\phi\dmd{\Psi_1}{4}{z}{2}{y}{2}+\dfrac{\sin4\phi}{2}\left(\dmd{\Psi_1}{4}{z}{}{y}{3}-\dmd{\Psi_1}{4}{z}{3}{y}{}\right)\right)=F_z(\psi_0,\theta_0),\label{psiz}
\end{eqnarray}
where $\Psi_1(z,y)=\psi_1(x-t,y)$ and $F_z$ is
\begin{equation} \label{Fz}
F_z=\frac{m_1+m_3}{2}+\frac{m_1-m_3}{2}\cos 2z+\frac{m_2}{2}\sin 2z,
\end{equation}
and the boundary conditions, \eqref{ep2bc1} and \eqref{ep2bc2}, are
\begingroup
\addtolength{\jot}{0.7em}
\begin{eqnarray}
\dpd{\Psi_1}{y}\bigg|_{y=0}&=&\frac{1}{2}\left((\alpha_1\alpha_2-\beta_1\beta_2)(1-\cos 2z)-(\alpha_1\beta_2-\alpha_2\beta_1)\sin 2z\right) ,\label{ep2bc1z}\\
\dpd{\Psi_1}{z}\bigg|_{y=0}&=&0.\label{ep2bc2z}
\end{eqnarray}
\endgroup

The periodic nature of the swimming sheet means $\Psi_1$ is also periodic in $z$ and so, upon taking the $z$-average of the system \eqref{psiz}-\eqref{ep2bc2z}, the $z$ derivatives disappear and the system becomes
\begin{equation}\label{psiz_av}
\left(1+\frac{\mu_2}{4}\sin^22\phi+\mu_3\right)\dpd[4]{\overline{\Psi}_1^z}{y}=\sum_{k=1}^{10}\frac{(M_1^{(k)}+M_3^{(k)})}{2}e^{\gamma_k y},
\end{equation}
\begingroup
\addtolength{\jot}{0.7em}
\begin{eqnarray}
\dpd{{\overline{\Psi}_1^z}}{y}\bigg|_{y=0}&=&\frac{1}{2}(\alpha_1\alpha_2-\beta_1\beta_2) ,\label{ep2bc1z_av}\\
\dpd{{{\overline{\Psi}_1^z}}}{z}\bigg|_{y=0}&=&0,\label{ep2bc2z_av}
\end{eqnarray}
\endgroup
where $\overline{\,\cdot\,}^z\equiv\frac{1}{2\pi}\int_{-\pi}^{\pi}\cdot\dif z$.

At this stage in the original paper an incorrect ansatz was assumed which neglected inhomogeneous terms. Hence we alter this ansatz to correctly determine the first order stream function $\Psi_1$ and thus the swimming velocity. Consider a complementary solution to the homogeneous problem and a particular integral satisfying the inhomogeneous portion; {\it i.e.}
\begin{equation}
\overline{\Psi}^z_1(y)=\overline{\Psi}^z_{C}(y)+\overline{\Psi}^z_{P}(y).
\end{equation}
For the homogeneous problem we have
\begin{equation} \label{hom}
\left(1+\frac{\mu_2}{4}\sin^22\phi+\mu_3\right)(\overline{\Psi}^z_{C})''''=0,
\end{equation}
with general solution $\overline{\Psi}^z_{C}(y)=A_3y^3+A_2y^2+A_1y+A_0$, where $'\equiv d/dy$.

The inhomogeneous problem is
\begin{equation}\label{inhom}
\left(1+\frac{\mu_2}{4}\sin^22\phi+\mu_3\right)(\overline{\Psi}^z_{P})''''=\sum_{k=1}^{10}\frac{(M_1^{(k)}+M_3^{(k)})}{2}e^{\gamma_k y},
\end{equation}
hence assume $\overline{\Psi}^z_{P}(y)$ takes the form
\begin{equation}
\overline{\Psi}^z_{P}=\sum_{k=1}^{10}P^{(k)}e^{\gamma_k y},
\end{equation}
where $P^{(k)}$ are constants to be determined. Substituting this form into \eqref{inhom} and rearranging for constants $P^{(k)}$, we find
\begin{equation}
P^{(k)}=\frac{M^{(k)}}{\left(1+\frac{\mu_2}{4}\sin^22\phi+\mu_3\right)\gamma_k^4},
\end{equation}
for each $k$ and where $M^{(k)}=(M_1^{(k)}+M_3^{(k)})/2$. Combining these two solutions,
\begin{equation} \label{psi_sol}
\overline{\Psi}^z_1=A_3y^3+A_2y^2+A_1y+A_0+\frac{1}{\left(1+\frac{\mu_2}{4}\sin^22\phi+\mu_3\right)}\sum_{k=1}^{10}\frac{M^{(k)}}{\gamma_k^4}e^{\gamma_ky}.
\end{equation}

The boundary conditions are used to determine the constants in equation \eqref{psi_sol}; for the velocity to be bounded we require $A_3=A_2=0$ and for the $z$-averaged problem the boundary condition \eqref{ep2bc1} becomes
\begin{equation}
\dpd{{\overline{\Psi}^z_1}}{y}\bigg|_{y=0}=\frac{1}{2}(\alpha_1\alpha_2-\beta_1\beta_2),
\end{equation}
which yields
\begin{equation}
A_1=\frac{1}{2}(\alpha_1\alpha_2-\beta_1\beta_2)+\dfrac{\mu_2(\cos4\phi+\cos2\phi)}{8(1+\frac{\mu_2}{4}\sin^22\phi+\mu_3)},
\end{equation}
and $A_0$ can be set to zero without loss of generality. Hence the full solution is
\begin{eqnarray}
\overline{\Psi}^z_1&=y\left[\dfrac{1}{2}(\alpha_1\alpha_2-\beta_1\beta_2)+\dfrac{\mu_2(\cos 4\phi+\cos2\phi)}{8(1+\frac{\mu_2}{4}\sin^22\phi+\mu_3)}\right] \nonumber \\
& \hspace{2.7cm} +\dfrac{1}{\left(1+\dfrac{\mu_2}{4}\sin^22\phi+\mu_3\right)}\sum_{k=1}^{10}\dfrac{M^{(k)}}{\gamma_k^4}e^{\gamma_ky}.
\end{eqnarray}
Due to the periodicity of the problem, the time and $z$-averages are identical, {\it i.e.} $\overline{U}^z=\overline{U}^t$. In the far field the mean swimming velocity is thus
\begin{equation} \label{swimvel}
\overline{U}^t=\frac{1}{2}(\alpha_1\alpha_2-\beta_1\beta_2)+\frac{\mu_2(\cos 4\phi+\cos2\phi)}{8(1+\frac{\mu_2}{4}\sin^22\phi+\mu_3)},
\end{equation}
where $\overline{\,\cdot\,}^t\equiv\frac{1}{2\pi}\int_0^{2\pi}\cdot\dif t$.
The second term in this solution was not included in our previous analysis. In comparison, the solution presented by Shi \& Powers is
\begin{equation} \label{SP_U}
U=\frac{1}{2}+\frac{\mu_2}{4}(\cos 4\phi + \cos 2\phi).
\end{equation}
Equations \eqref{swimvel} and \eqref{SP_U} agree in the limit $\mu_2\rightarrow0$ and $\mu_3=0$, modulo a factor of two introduced in the analysis by \citeauthor{shi2017swimming} (which can be absorbed into $\mu_2$).

\subsection{Results} \label{no_mu1_Results}
We present the new results for a range of parameter values and compare them with those produced by Shi \& Powers.

\begin{figure}
\centering
\includegraphics[scale=0.75]{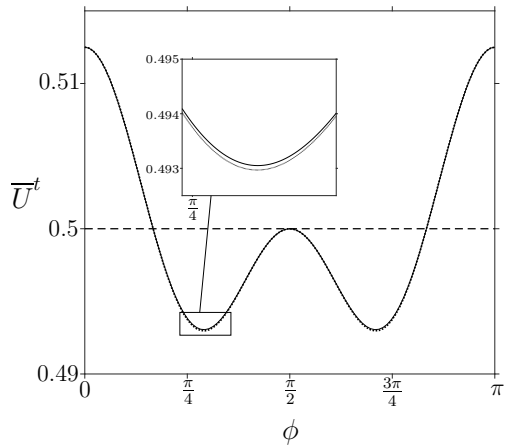}
\caption{Mean swimming velocity comparison for passive transversely-isotropic media where $\mu_2=0.05$ and $\mu_3=0$. Three different results are compared: the incorrect calculation from \citet{cupples2017viscous} (dashed line), the corrected calculation (solid line) and the solution provided by \citet{shi2017swimming} (dotted line). A magnified view of the first minimum in this figure has been included.}
\label{fig:U_comp}
\end{figure}

Firstly we make a direct comparison with the work by Shi \& Powers in figure \ref{fig:U_comp}; our original calculation, $\overline{U}^t=(\alpha_1\alpha_2-\beta_1\beta_2)/2$ (equation (3.36) in \citeauthor{cupples2017viscous} \citeyear{cupples2017viscous}) is plotted as the dashed line, the corrected solution is the solid line and the Shi \& Powers result is shown by the dotted line. In addition to $\mu_1=0$, the anisotropic shear viscosity $\mu_3$ is set to zero and $\mu_2=0.05$. It is immediately seen that the inclusion of the extra term has altered the mean swimming velocity and introduced a dependence on the initial orientation angle $\phi$. Aside from the minimum values of the mean swimming velocity, the corrected solution agrees well with the work from Shi \& Powers; this small difference is due to the $1/(1+\mu_2\sin^22\phi/4+\mu_3)$ multiplying the second term in \eqref{swimvel}, as can be seen in the magnified view in figure \ref{fig:U_comp}.

\begin{figure}
\centering
\includegraphics[scale=0.75]{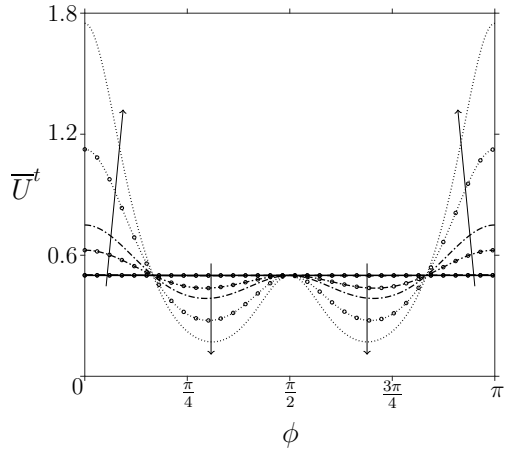}
\caption{Corrected mean swimming velocity for small $\mu_2$ and $\mu_3$. Four $\mu_2$ values are chosen, $\mu_2=0$ (solid lines), $\mu_2=0.01$ (dashed lines), $\mu_2=1$ (dot-dashed lines) and $\mu_2=5$ (dotted lines). Two $\mu_3$ values are selected, $\mu_3=0$ and $\mu_3=1$ (circle markers).}
\label{fig:U_small}
\end{figure}

Next we consider a larger range of $\mu_2$ and $\mu_3$ and compare the mean swimming velocity. First consider small $\mu_2$ and $\mu_3$ (figure \ref{fig:U_small}). Increasing $\mu_2$ dominates the impact of the initial orientation angle on the mean swimming velocity, and the anisotropic shear viscosity works to collapse the results back towards the Newtonian value; when both parameters are zero we return to the Newtonian solution as expected. For very small $\mu_2$ (dashed lines which are not seen) the variation from the Newtonian solution is very small.

\begin{figure}
\centering
\includegraphics[scale=0.65]{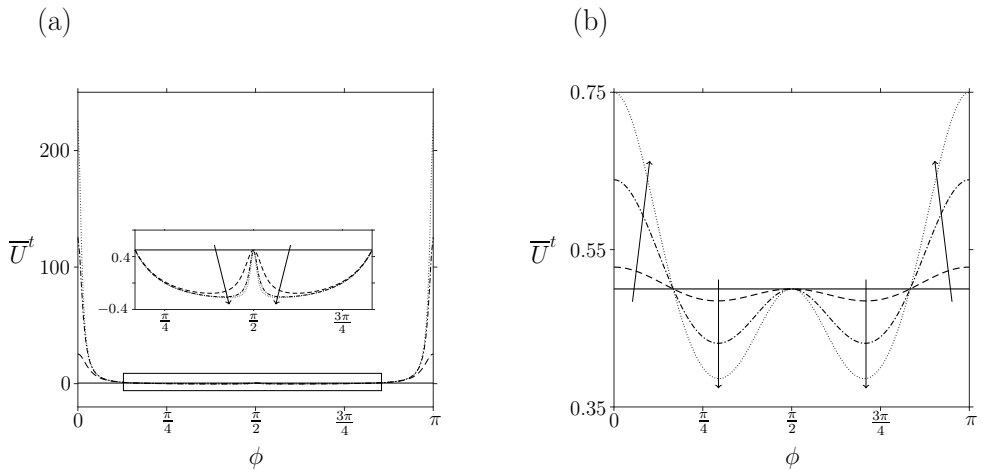}
\caption{Corrected mean swimming velocity for large $\mu_2$. (a) $\mu_3=0$ and (b) $\mu_3=900$. Four choices for $\mu_2$ are compared: $\mu_2=0$ (solid lines), $\mu_2=100$ (dashed lines), $\mu_2=500$ (dot-dashed lines) and $\mu_2=900$ (dotted lines). Figure (a) contains a magnified view of the middle section of the results.}
\label{fig:U_large}
\end{figure}

Finally we investigate the impact when both $\mu_2$ and $\mu_3$ may take on large values. Here we have separated the results into two cases; when $\mu_3=0$ (figure \ref{fig:U_large}a) and when $\mu_3=900$ (figure \ref{fig:U_large}b). When $\mu_3=0$, the mean swimming velocity takes on large values near $\phi=0$ and $\phi=\pi$; these sharp peaks are consistent with the results in \citep{cupples2017viscous} occurring when one parameter was much larger than the others. Away from these regions, the mean swimming velocity takes on values similar to those presented in figures \ref{fig:U_comp} and \ref{fig:U_small}. When both parameters are large (figure \ref{fig:U_large}b), the mean swimming velocity reduces in comparison to figure \ref{fig:U_large}a. The shape of the $\phi$-$\overline{U}^t$ curve is similar as the anisotropic parameters are varied, only the magnitude changes.

\section{Mean swimming velocity in active media} \label{mu1_sol}

Next consider active transversely-isotropic media, where $\mu_1\neq0$. The equations governing the flow and orientation are given by \eqref{gov1} and \eqref{gov2} respectively. Due to the time derivatives that force the evolution of orientation (equation \eqref{gov2}) we can no longer seek a solution depending on $z=x-t$ and instead look at an $x$-average of the coupled system.

\subsection{Corrected solution} \label{corrected_mu1}
Based on the geometry of the problem, $\psi_1$ and $\theta_1$ will be periodic in $x$. Hence, an $x$-average is taken,
\begin{eqnarray}
\left(1+\frac{\mu_2}{4}\sin^22\phi+\mu_3\right)\dpd[4]{{\overline{\psi}^x_1}}{y}+\mu_1\cos2\phi\dpd[2]{{\overline{\theta}^x_1}}{y}&=&\frac{m_1+m_3}{2}, \label{av1}\\
\dpd{{\overline{\theta}^x_1}}{t}+\sin^2\phi\dpd[2]{{\overline{\psi}^x_1}}{y}&=&\frac{n_1+n_3}{2}, \label{av2}
\end{eqnarray}
where $\overline{\,\cdot\,}^x\equiv\frac{1}{2\pi}\int_0^{2\pi}\cdot\dif x$. Equation \eqref{av1} can be directly integrated twice with respect to $y$,
\begin{equation} \label{av1_int}
\left(1+\frac{\mu_2}{4}\sin^22\phi+\mu_3\right)\dpd[2]{{\overline{\psi}^x_1}}{y}+\mu_1\cos2\phi\,{\overline{\theta}^x_1}=\sum_{k=1}^{10}\frac{M^{(k)}e^{\gamma_k y}}{\gamma_k^2}+B_0(t)y+B_1(t),
\end{equation}
and substituted into \eqref{av2} to give
\begin{eqnarray}
\dpd{{\overline{\theta}^x_1}}{t}&-&\frac{\mu_1\cos2\phi\sin^2\phi}{1+\frac{\mu_2}{4}\sin^22\phi+\mu_3}\overline{\theta}^x_1=\sum_{k=1}^{10}N^{(k)}e^{\gamma_k y} \nonumber \\
& & \hspace{1cm}-\frac{\sin^2\phi}{1+\frac{\mu_2}{4}\sin^22\phi+\mu_3}\left(\sum_{k=1}^{10}\frac{M^{(k)}e^{\gamma_k y}}{\gamma_k^2}+B_0(t) y+B_1(t)\right),  \label{dth_dt}
\end{eqnarray}
where $B_0(t)$ and $B_1(t)$ are functions of time to be determined. To simplify the following calculations, the functions are written in the form $B_0(t)=\dot{F}_0(t)\exp(\mu_1\Gamma t)$ and $B_1(t)=\dot{F}_1(t)\exp(\mu_1\Gamma t)$, where the dot notation represents a time derivative, $F_0(t)$ and $F_1(t)$ are functions of time to be determined and
\begin{equation} \label{gamma}
\Gamma=\frac{\cos2\phi\sin^2\phi}{1+\frac{\mu_2}{4}\sin^22\phi+\mu_3}.
\end{equation}
Then, equation \eqref{dth_dt} can be solved via an integrating factor to give
\begin{equation}
\overline{\theta}_1=\frac{f(y)}{\mu_1\Gamma}-\frac{\Gamma}{\cos2\phi}(F_0(t) y+F_1(t))e^{\mu_1\Gamma t} + c(y)e^{\mu_1\Gamma t},
\end{equation}
where
\begin{equation}
f(y)=-\sum_{k=1}^{10}N^{(k)}e^{\gamma_k y}+\frac{\Gamma}{\cos2\phi}\sum_{k=1}^{10}\frac{M^{(k)}e^{\gamma_k y}}{\gamma_k^2},
\end{equation}
and $c(y)$ is a function to be determined. The full solution is detailed in appendix \ref{AppB}. Since the fibres have initial orientation $\phi$, the initial condition for the angle is $\overline{\theta}^x_1(x,y,0)=0$ and so
\begin{equation}
c(y)=-\frac{f(y)}{\mu_1\Gamma}+\frac{\Gamma}{\cos2\phi}(F_0^0y+F_1^0),
\end{equation}
where $F_j^0=F_j(0)$. The solution is thus
\begin{equation}
\overline{\theta}^x_1=\frac{f(y)}{\mu_1\Gamma}(1-e^{\mu_1\Gamma t})-\frac{\Gamma}{\cos2\phi}(B_2(t)y+B_3(t))e^{\mu_1\Gamma t},
\end{equation}
for $B_2(t)=F_0(t)-F_0^0$, $B_3(t)=F_1(t)-F_1^0$.

The form for $\overline{\theta}^x_1$ can now be substituted back into equation \eqref{av1_int},
\begin{eqnarray}
& &(1+\frac{\mu_2}{4}\sin^22\phi+\mu_3)\dpd[2]{{\overline{\psi}^x_1}}{y}=\frac{\cos2\phi}{\Gamma}(1-e^{\mu_1\Gamma t})\sum_{k=1}^{10}N^{(k)}e^{\gamma_k y} \nonumber \\
& &\hspace{0.8cm} +e^{\mu_1\Gamma t}\sum_{k=1}^{10}\frac{M^{(k)}e^{\gamma_k y}}{\gamma_k^2}+\mu_1\Gamma(B_2(t)y+B_3(t))e^{\mu_1\Gamma t}+B_0(t)y+B_1(t),
\end{eqnarray}
which can then be directly integrated with respect to $y$ to obtain
\begin{eqnarray}
& &(1+\frac{\mu_2}{4}\sin^22\phi+\mu_3)\overline{\psi}^x_1=\frac{\cos2\phi}{\Gamma}(1-e^{\mu_1\Gamma t})\sum_{k=1}^{10}\frac{N^{(k)}e^{\gamma_k y}}{\gamma_k^2} \nonumber \\
& &\hspace{3cm} +e^{\mu_1\Gamma t}\sum_{k=1}^{10}\frac{M^{(k)}e^{\gamma_k y}}{\gamma_k^4} + \left(B_0(t)+\mu_1\Gamma B_2(t)e^{\mu_1\Gamma t}\right)\frac{y^3}{6} \nonumber \\
& & \hspace{3cm} +\left(B_1(t)+\mu_1\Gamma B_3(t)e^{\mu_1\Gamma t}\right)\frac{y^2}{2}+B_4(t) y+B_5(t).
\end{eqnarray}
To determine the functions of integration, reconsider the boundary condition
\begin{equation} \label{bc_mu1_ref}
\dpd{{\overline{\psi}^x_1}}{y}\bigg|_{y=0}=\frac{1}{2}(\alpha_1\alpha_2-\beta_1\beta_2),
\end{equation}
and note that since the velocity must remain bounded in the far field we require $B_0(t)+\mu_1\Gamma B_2(t)e^{\mu_1\Gamma t}=0$ and $B_1(t)+\mu_1\Gamma B_3(t)e^{\mu_1\Gamma t}=0$.  It can be shown that this is equivalent to $B_0(t)=B_1(t)=B_2(t)=B_3(t)=0$ (see appendix \ref{constants}). Since $B_5$ has no impact on the velocity it can, without loss of generality, be set to zero. The final function $B_4(t)$ is determined from equation \eqref{bc_mu1_ref} as
\begin{eqnarray}
B_4(t)&=&\frac{1}{2}(\alpha_1\alpha_2-\beta_1\beta_2)+\frac{1}{\sin^2\phi}(e^{\mu_1\Gamma t}-1)\sum_{k=1}^{10}\frac{N^{(k)}}{\gamma_k^2} \nonumber \\
& &\hspace{1cm} -\frac{e^{\mu_1\Gamma t}}{1+\frac{\mu_2}{4}\sin^22\phi+\mu_3}\sum_{k=1}^{10}\frac{M^{(k)}}{\gamma_k^4}.
\end{eqnarray}

Hence, the solutions $\overline{\psi}^x_1$ and $\overline{\theta}^x_1$ are given by
\begin{eqnarray}
\overline{\psi}^x_1&=&y\left[\frac{1}{2}(\alpha_1\alpha_2-\beta_1\beta_2)\right.\nonumber \\
& &\hspace{1cm} \left.+\frac{e^{\mu_1\Gamma t}-1}{\sin^2\phi}\sum_{k=1}^{10}\frac{N^{(k)}}{\gamma_k}-\frac{e^{\mu_1\Gamma t}}{1+\frac{\mu_2}{4}\sin^22\phi+\mu_3}\sum_{k=1}^{10}\frac{M^{(k)}}{\gamma_k^3}\right] \nonumber\\
& & \hspace{1cm}-\frac{e^{\mu_1\Gamma t}-1}{\sin^2\phi}\sum_{k=1}^{10}\frac{N^{(k)}e^{\gamma_k y}}{\gamma_k^2}+\frac{e^{\mu_1\Gamma t}}{1+\frac{\mu_2}{4}\sin^22\phi+\mu_3}\sum_{k=1}^{10}\frac{M^{(k)}e^{\gamma_k y}}{\gamma_k^4}, \label{psi_mu1_sol}\\
\overline{\theta}^x_1&=&\left[\frac{1+\frac{\mu_2}{4}\sin^22\phi+\mu_3}{\mu_1\cos2\phi\sin\phi^2}\sum_{k=1}^{10}N^{(k)}e^{\gamma_k y}\right. \nonumber \\
& & \hspace{4cm}\left. -\frac{1}{\mu_1\cos2\phi}\sum_{k=1}^{10}\frac{M{(k)}e^{\gamma_k y}}{\gamma_k^2}\right](e^{\mu_1\Gamma t}-1),  \label{th_mu1_sol}
\end{eqnarray}
where the swimming velocity is given at far field as
\begin{equation} \label{U_mu1_sol}
\overline{U}^x=e^{\mu_1\Gamma t}\left[\frac{1}{\sin^2\phi}\sum_{k=1}^{10}\frac{N^{(k)}}{\gamma_k}-\frac{1}{1+\frac{\mu_2}{4}\sin^22\phi+\mu_3}\sum_{k=1}^{10}\frac{M^{(k)}}{\gamma_k^3}\right].
\end{equation}

\subsection{Comments}
Equation \eqref{U_mu1_sol} will be valid only when $\mu_1\Gamma\leq0$ or for very short time scales. The sign of $\mu_1\Gamma$ is determined by $\mu_1\cos2\phi$; for `puller' type behaviour, where $\mu_1$ is positive, the solution is valid only for $\pi/4\leq\phi\leq3\pi/4$ and these exponential terms decay with time. This however leads to a steady-state swimming velocity $\overline{U}^x=0$ and so the active properties of the fluid halt any propulsion. For `pusher' type behaviour, where $\mu_1$ is negative, this validity is for $0\leq\phi\leq\pi/4$ and the same result for the swimming velocity is obtained.

Outside this region, the solution for $\overline{\theta}^x_1$ and further the swimming velocity $\overline{U}^x$ grow exponentially and hence will not be valid in the perturbation expansion currently considered. To fully understand microscopic propulsion in active transversely-isotropic media, it will be necessary to consider a numerical solution to the full swimming problem.

\section{Discussion}
A corrigendum to `Viscous propulsion in active transversely-isotropic media' has been described, prompted by \citet{shi2017swimming} who investigated propulsion in nematic liquid crystals and discovered a discrepancy between the two models in a common limit. The corrected swimming velocity was calculated for a passive fluid, from which it was found that the extra terms introduce a dependence of the mean swimming velocity on the initial orientation angle. By setting $\mu_2$ to be small and $\mu_3=0$ our corrected result agrees with \citet{shi2017swimming} in the common limit.

The corrected swimming velocity was then compared for a range of $\mu_2$ and $\mu_3$. The effects of the initial orientation angle on $\overline{U}^t$ were increased by increasing the anisotropic extensional viscosity and larger anisotropic shear viscosities reduce the effect of the initial orientation angle. Further, when one parameter is large and the other small, rapid changes in the swimming velocity and a reversal in the swimming direction ({\it i.e.} negative swimming velocity) were seen; a result seen consistent with the mean rate of working found in \citet{cupples2017viscous}.

Finally a solution for the swimming velocity in active media ($\mu_1\neq0$) was sought. Periodicity in $x$ was imposed for the stream function and evolution of orientation angle due to the problem geometry; this observation simplified the calculations required. The coupled equations were solved to determine the first order evolution of orientation and the swimming velocity. The swimming velocity varied exponentially in time, with the sign of the exponent dependent on $\mu_1$ and the initial orientation angle. Thus the expansion is valid only for very short time periods, or for specific $\mu_1$ and initial orientation angles where the exponent is negative; in these cases the active properties appear to halt propulsion. Setting $\mu_1=0$ returned the result for the passive case. A topic of significant interest for future work is to investigate a fully numerical solution to the swimming problem in active transversely-isotropic media.

\section{Acknowledgements}
This work was supported by a  Biotechnology  and  Biological  Sciences  Research  Council (BBSRC)  Industrial CASE Studentship (BB/L015587/1). The authors acknowledge Profs.\ J.\ Shi and T.R.\ Powers for identifying the discrepancy between our respective papers.

\bigskip

\appendix

\section{Inhomogeneous terms in $\mathcal{O}(\varepsilon^2)$ governing equations} \label{AppA}
The right hand side of equations \eqref{gov1} and \eqref{gov2} are
\begin{eqnarray}
& &F(\psi_0,\theta_0)=\mu_1\left[2\sin2\phi\left(\theta_0\left(\dpd[2]{\theta_0}{y}-\dpd[2]{\theta_0}{x}\right)+\left(\dpd{\theta_0}{y}\right)^2-\left(\dpd{\theta_0}{x}\right)^2\right)\right. \nonumber \\
& &\left. +4\cos2\phi\left(\dpd{\theta_0}{x}\dpd{\theta_0}{y}+\theta_0\dmd{\theta_0}{2}{x}{}{y}{}\right) \right]+\mu_2\left[\sin4\phi\left(2\dmd{\theta_0}{2}{x}{}{y}{}\dmd{\psi_0}{2}{x}{}{y}{} \right.\right.\nonumber \\
& &\left.\left. -\frac{\theta_0}{2}\left(\dpd[4]{\psi_0}{x}-6\dmd{\psi_0}{4}{x}{2}{y}{2}+\dpd[4]{\psi_0}{y}\right) +\frac{1}{2}\left(\dpd[2]{\theta_0}{x}-\dpd[2]{\theta_0}{y}\right)\left(\dpd[2]{\psi_0}{y}-\dpd[2]{\psi_0}{x}\right)\right.\right. \nonumber\\
& &\qquad \left.\left.-\dpd{\theta_0}{y}\left(\dpd[3]{\psi_0}{y}-3\dmd{\psi_0}{3}{x}{2}{y}{}\right)+\dpd{\theta_0}{x}\left(3\dmd{\psi_0}{3}{x}{}{y}{2}-\dpd[3]{\psi_0}{x}\right)\right)\right. \nonumber\\
& &\qquad \left.+\cos4\phi\left(2\theta_0\left(\dmd{\psi_0}{4}{x}{3}{y}{}-\dmd{\psi_0}{4}{x}{}{y}{3}\right)+\left(\dpd[2]{\theta_0}{x}-\dpd[2]{\theta_0}{y}\right)\dmd{\psi_0}{2}{x}{}{y}{} \right.\right.\nonumber  \\
& &\qquad \left.\left. -\dpd{\theta_0}{x}\left(\dpd[3]{\psi_0}{y}-3\dmd{\psi_0}{3}{x}{2}{y}{}\right)-\dpd{\theta_0}{y}\left(3\dmd{\psi_0}{3}{x}{}{y}{2}-\dpd[3]{\psi_0}{x}\right)\right.\right. \nonumber \\
& &\hspace{5cm} \left.\left. -\dmd{\theta_0}{2}{x}{}{y}{}\left(\dpd[2]{\psi_0}{y}-\dpd[2]{\psi_0}{x}\right)\right)\right], \label{extraF}
\end{eqnarray}
\begin{eqnarray}
G(\psi_0,\theta_0)&=&\dpd{\psi_0}{x}\dpd{\theta_0}{y}-\dpd{\psi_0}{y}\dpd{\theta_0}{x} \nonumber \\
& &-\theta_0\left(2\cos2\phi\dmd{\psi_0}{2}{x}{}{y}{}+\sin2\phi\left(\dpd[2]{\psi_0}{y}-\dpd[2]{\psi}{x}\right)\right).\label{extraG}
\end{eqnarray}

\section{Evolution of orientation in an active suspension}\label{AppB}
To determine the swimming velocity and evolution of orientation from system \eqref{av1}-\eqref{av2}, firstly \eqref{av1} is integrated twice (equation \eqref{av1_int}) and substituted into \eqref{av2} resulting in equation \eqref{dth_dt},
\begin{eqnarray}
\dpd{{\overline{\theta}^x_1}}{t}&-&\frac{\mu_1\cos2\phi\sin^2\phi}{1+\frac{\mu_2}{4}\sin^22\phi+\mu_3}\overline{\theta}^x_1=\sum_{k=1}^{10}N^{(k)}e^{\gamma_k y} \nonumber \\
& & \hspace{1cm}-\frac{\sin^2\phi}{1+\frac{\mu_2}{4}\sin^22\phi+\mu_3}\left(\sum_{k=1}^{10}\frac{M^{(k)}e^{\gamma_k y}}{\gamma_k^2}+B_0(t) y+B_1(t)\right).
\end{eqnarray}
We set $B_0(t)=F'_0(t)\exp(\mu_1\Gamma t)$ and $B_1(t)=F'_1(t)\exp(\mu_1\Gamma t)$, where $\Gamma$ is given by \eqref{gamma}, to simplify the following calculations.

Introduce an integrating factor such that
\begin{eqnarray}
\dpd{}{t}(e^{-\mu_1\Gamma t} \overline{\theta}^x)&=& e^{-\mu_1\Gamma t}\sum_{k=1}^{10}N^{(k)}e^{\gamma_k y} \nonumber \\
&-& \frac{\Gamma}{\cos2\phi}\left(e^{-\mu_1\Gamma t} \sum_{k=1}^{10}\frac{M^{(k)}e^{\gamma_k y}}{\gamma_k^2}+F_0'y+F_1'\right).
\end{eqnarray}
Next integrate with respect to $t$ and rearrange to give
\begin{eqnarray}
\overline{\theta}^x&=& -\frac{1}{\mu_1\Gamma}\sum_{k=1}^{10}N^{(k)}e^{\gamma_k y}+ \frac{\Gamma}{\cos2\phi}\left(\frac{1}{\mu_1\Gamma} \sum_{k=1}^{10}\frac{M^{(k)}e^{\gamma_k y}}{\gamma_k^2}\right) \nonumber \\
& & \hspace{2cm} -\frac{\Gamma}{\cos2\phi}\left(F_0 y+F_1\right)e^{\mu_1\Gamma t}+c(y) e^{\mu_1\Gamma t},
\end{eqnarray}
which is simplified to
\begin{equation}
\overline{\theta}^x_1=\frac{f(y)}{\mu_1\Gamma}-\frac{\Gamma}{\cos2\phi}(F_0 y+F_1)e^{\mu_1\Gamma t} + c(y)e^{\mu_1\Gamma t},
\end{equation}
for function $c(y)$ determined via boundary condition \eqref{bc_mu1_ref} in section \ref{corrected_mu1}.

\section{Finding constants of integration}\label{constants}

From (\ref{bc_mu1_ref}) we require $B_0(t)+\mu_1\Gamma B_2(t)e^{\mu_1\Gamma t}=0$, and $B_1(t)+\mu_1\Gamma B_3(t)e^{\mu_1\Gamma t}=0$.  We here show that this is equivalent to $B_0(t)=B_1(t)=B_2(t)=B_3(t)=0$.

Recall the forms $B_0(t)=\dot{F}_0(t)e^{\mu_1\Gamma t}$ and $B_2(t)=F_0(t)-F_0^0$, and similarly for $B_1$ and $B_3$ respectively. Substituting this form into the expressions satisfying the far-field condition we have
\begin{eqnarray}
\dot{F}_0(t)e^{\mu_1\Gamma t}+\mu_1\Gamma e^{\mu_1\Gamma t}(F_0(t)-F_0^0)=0, \label{farfieldcheck}\\
\dot{F}_1(t)e^{\mu_1\Gamma t}+\mu_1\Gamma e^{\mu_1\Gamma t}(F_1(t)-F_1^0)=0.
\end{eqnarray}
Since both equations have the same form, we illustrate the solution for equation \eqref{farfieldcheck} only; rearrange and integrate with respect to $t$ to obtain
\begin{equation}
F_0(t)=F_0^0+De^{-\mu_1\Gamma t},
\end{equation}
for some constant $D$. As $F_0(0)=F_0^0$, we find $D=0$ and hence $F_0$ is constant in time; this then enforces $B_0(t)=B_2(t)=0$.

\end{document}